\documentclass[conference]{IEEEtran}
\IEEEoverridecommandlockouts
\usepackage{cite}
\usepackage{amsmath,amssymb,amsfonts}
\usepackage{algorithmic}
\usepackage{graphicx}
\usepackage{textcomp}
\usepackage{xcolor}
\def\BibTeX{{\rm B\kern-.05em{\sc i\kern-.025em b}\kern-.08em
    T\kern-.1667em\lower.7ex\hbox{E}\kern-.125emX}}
\begin{document}

\title{Tamper-Evident Pairing\\
}

\author{\IEEEauthorblockN{Aleksandar Manev}
\IEEEauthorblockA{
\textit{Technical University of Munich}\\
Munich, Germany \\
aleksandar.manev@tum.de}
}

\maketitle
\begin{abstract}
Establishing a secure connection between wireless devices has become significantly important with the increasing number of Wi-Fi products coming to the market. In order to provide an easy and secure pairing standard, the Wi-Fi Alliance has designed the Wi-Fi Protected Setup. Push-Button Configuration (PBC) is part of this standard and is especially useful for pairing devices with physical limitations. However, PBC is proven to be vulnerable to man-in-the-middle (MITM) attacks.
Tamper-Evident Pairing (TEP) is an improvement of the PBC standard, which aims to fix the MITM vulnerability without interfering the useful properties of PBC. It relies on the Tamper-Evident Announcement (TEA), which guarantees that an adversary can neither tamper a transmitted message without being detected, nor hide the fact that the message has been sent. 
The security properties of TEP were proven manually by its authors and tested with the \textsc{Uppaal} and \textsc{Spin} model checkers. During the \textsc{Uppaal} model checking, no vulnerabilities were found. However, the \textsc{Spin} model revealed a case, in which the TEP's security is not guaranteed.
In this paper, we first provide a comprehensive overview of the TEP protocol, including all information needed to understand how it works. Furthermore, we summarize the security checks performed on it, give the circumstances, under which it is no longer resistant to MITM attacks and explain the reasons why they could not be revealed with the first model. Nevertheless, future work is required to gain full certainty of the TEP's security before applying it in the industry. 
\end{abstract}


\section{Introduction}\label{sec1}
Nowadays, it is hard to imagine living without access to the Internet. People all over the world rely on it for everything from entertainment to work and achieving their goals. Thanks to the wireless technology, it is now possible to connect to the Internet even remotely. The Wi-Fi Standard has become an inseparable part of people's lives and significantly important for keeping them connected with each other. And not only people communicate over wireless networks, but also more and more devices.

Such a high importance for the daily life implies a lot of information and personal data being transmitted over the wireless networks. This makes them an attractive target for hackers who try to exploit security flaws to gain access to private data. As we continue into a future, in which everything from our mobile phone to our home operates using a wireless network connection, it is becoming increasingly important to understand how to keep our wireless network safe and secure.

In order to connect devices securely in publicly accessible networks, pairing protocols are developed, which aim to ensure security properties such as authentication, confidentiality, and integrity. A pairing protocol has to prevent malicious users from intercepting and tampering existing messages as well as from sending their own over the network. However, the inherently shared nature of the wireless medium makes the secure wireless communication a challenging problem. Furthermore, if the security setup is too complex, ordinary users tend to skip doing it, which makes them an easy target for attackers. Therefore, keeping the user's involvement as simple as possible is a major requirement while designing a pairing protocol.

Another challenge for the security is that Wi-Fi devices often have limitations on their size and form due to their purpose or design. Such devices are, for instance, medical devices, home appliances, printers, cameras, motion and light sensors, etc. Thus, it is sometimes impossible to support an interface for entering a password (e.g., a display or a keyboard) to authenticate the user. 

In order to meet the needs, the Wi-Fi Alliance proposed a possible solution by introducing the Wi-Fi Protected Setup (WPS) \cite{alliancewi} in 2006. The WPS was designed to provide a standard for easy establishment of a secure Wi-Fi connection between a wireless device (with possibly limited interface) and a wireless access point in home and small office environments. The standard provides several familiar to the costumers options for configuring the security settings, among which by entering a key (the PIN method) and by pushing a button (Push-Button Configuration (PBC)). WPS successfully eased the setup of security-enabled Wi-Fi networks and solved the problem with the limited interface. Unfortunately, a major security gap of the PIN method was found by S. Viehb{\"o}ck \cite{viehbock2011brute} in 2011, which proved that it is vulnerable to brute-force attacks (see \ref{sec2}).

In order to establish a secure connection using the PBC method, the user needs to just press a button on each device within a certain time frame (usually 120 s). The devices exchange their Diffie-Hellman public keys \cite{journals/tit/DiffieH76}, which are used to protect their future communication. Diffie-Hellman's key exchange protocol guarantees protection against passive adversaries, i.e., adversaries, snooping on the wireless medium and trying to obtain key exchange messages. However, since these messages are not authenticated in any way, an adversary could impersonate each device to the other and convince both devices to pair via him. This makes PBC vulnerable to active man-in-the-middle (MITM) attacks.

Wi-Fi connection is widely used nowadays in very critical fields, e.g., in medical devices that transmit a patient's vital signals and home's security protection. Once connected to a wireless access point, a device gets not only Internet connectivity, but also access to all the data shared on the network and the opportunity to manipulate it. Therefore, using PBC could lead to letting an adversary in the wireless network, and thus to severe consequences. 

The PBC security issue could not be solved immediately due to the restrictions that such a solution would bring with itself. In order to prevent a MITM attack, both devices have to be authenticated, which would require that they support an out-of-band communication channel \cite{kainda2009usability}. This is a medium, different from the one used for transmitting normal data, e.g., a camera for facial recognition, a device with fingerprint recognition, etc. However, supporting such an out-of-band communication channel can be unallowable due to costs and imposed constraints on some Wi-Fi devices. Thus, it is difficult for the industry to adopt it as a solution.

To protect key establishment against MITM attacks without authentication through out-of-band channels, in 2011, Gollakota et al. presented a method, called Tamper-Evident Pairing (TEP) \cite{266516}, that provides simple and secure Wi-Fi \textit{in-band} pairing. TEP is based on PBC and introduces a new primitive – the Tamper-Evident Announcement (TEA). Furthermore, TEP can also be implemented on existing PBC and Wi-Fi devices without any hardware changes needed. The essence of the proposed method is that the IEEE 802.11 (see \ref{sec3.2}) devices are not only able to transmit data, but also to sense the medium to detect whether or not information is being transmitted. Based on the assumption that an adversary can only change or corrupt data on the medium, but not completely remove it, Gollakota et al. manually proved the correctness of TEP.

In general, there are multiple approaches to analyze security protocols. A protocol should work for arbitrarily many runs in all possible scenarios and in the presence of one or more adversaries. This makes analysis by hand difficult and has been a motivation for the development of specialized tools that address all these problems. The two general ways are model checking and theorem proving. 
In model checking, a property is investigated for a finite number of states, whereas in theorem proving, all possible states are considered. Therefore, the absence of errors in model checking does not imply correctness of a protocol. However, it is often useful to reveal vulnerabilities before constructing a proof with a theorem prover, which normally would require a lot of effort and time. An example for theorem proving tools for automatically analyzing security properties is Prototype Verification System \cite{owre1992pvs}. General purpose model checkers, such as \textsc{Spin} \cite{holzmann1997model}, \textsc{Uppaal} \cite{bengtsson1995uppaal} and \textsc{NuSMV} \cite{cimatti1999nusmv}, have also been successfully applied to verify desired properties of protocol specifications \cite{Basin2018}.

In 2012, M. Drijvers et al. \cite{drijvers2012model} used the model checker \textsc{Uppaal} to analyze TEP. First, the TEA was modeled. Its desired properties were then translated into \textsc{Uppaal}-queries and tested on the model. After all desired properties were shown to hold, TEP was implemented as a separate model that uses the results of the TEA model. No vulnerabilities were found.

One year later, in 2013, R. Kersten et al. \cite{kersten2013using} investigated TEP by using another model checker - \textsc{Spin}. In the original TEP specification, some parameters were used that were not fully specified by the authors. Kersten et al. proved that the values of these parameters could be critical for the security of the protocol by revealing a vulnerability for a specific scenario. Moreover, the appropriate values for the parameters that can break the resistance of TEP against MITM attacks were given. 

In this paper, we will introduce the possible reasons for not revealing the vulnerabilities of TEP at first. We will also include all of the needed information to fully understand how TEP works and why it would be useful to be applied in the industry. Nevertheless, the edge cases for the unspecified parameters that were revealed to cause a vulnerability should be at least avoided or precisely specified. Moreover, before applying TEP to the existing Wi-Fi devices, a formal verification of the security properties of the protocol would be needed.

\section{Related Work}\label{sec2}
The security properties of the Wi-Fi Protected Setup (WPS) have been analyzed a lot. In 2011, S. Viehböck found a massive vulnerability of the PIN method, provided by WPS \cite{viehbock2011brute}.
In order to pair two devices, an 8-digit authentication code (PIN) is displayed to the user on the first device, which then has to be entered in on the second device. However, this method is vulnerable to brute-force attacks. The {$8^{th}$} digit of the PIN appeared to be a checksum of the first seven (i.e., there are only 7 digits to determine and a brute-force attack would require max. $10^7$ attempts). Furthermore, Viehböck discovered that the PIN is verified in two steps and from the device's response to a connection attempt, an adversary can discover in which half of the PIN is the mistake. The adversary could thus first brute-force the first half ($10^4$ attempts at most), and then the second half ($10^3$ attempts at most). A successful brute-force attack would require only $10^4 + 10^3 = 11000$ attempts instead of $10^7 = 10 000 000$, which is feasible. In response to the found vulnerability, it was recommended for the users to disable the WPS feature on their wireless access points.

A comprehensive overview of different device pairing mechanisms is given in \cite{6687314}. The survey presents numerous existing protocols that, similar to PBC, use unauthenticated Diffie-Hellman key exchange messages, and are thus vulnerable to MITM attacks. The authors focus on the solution via out-of-band communication channels in the authentication procedure and show how the involvement of the user can be used to mitigate the MITM problem. Moreover, the survey presents protocols that allow secure group device pairing.

Bluetooth is another popular standard that realizes wireless communication between devices. Similar to the WPS, the Bluetooth's Secure Simple Pairing (SSP) aims to make it accessible for ordinary users by supporting multiple setup mechanisms. A thorough security and usability analysis of both WPS and SSP is provided by \cite{kuo2007low}. Successfully mounted MITM attacks in SSP are presented in \cite{4537388}, \cite{4679061} and \cite{5374082}, along with proposed countermeasures for security improvements.

\section{Background Information}\label{sec3}
We will introduce the theoretical concepts that are required to fully understand the Tamper-Evident Pairing. We start by describing the Diffie-Hellman Key Exchange as this is used in both PBC and TEP. Then the main concepts of IEEE 802.11 are provided, followed by a detailed description of PBC, on which TEP is based, and the assumed attacker model.

\subsection{Diffie-Hellman Key Exchange}\label{sec3.1}
Diffie-Hellman Key Exchange protocol \cite{journals/tit/DiffieH76} is a method for securely exchanging cryptographic keys over a public channel. It aims to establish a shared key between two parties that can be used for future data exchange over the public network. The protocol uses the multiplicative group of integers modulo $p$, where $p$ is prime, and $g$ primitive root modulo $p$, i.e., for every integer $a$ co-prime to $p$, there is some integer $k$ for which $g^k \equiv a \mod p$. These two values are chosen in this way to ensure that the resulting shared secret can take on any value between $1$ and $p - 1$.

Assuming Alice and Bob are the two parties, the process works as follows:
\begin{enumerate}
    \item Alice and Bob publicly agree on a non-secret prime numbers $p$ and $g$.
    \item Alice chooses a secret integer $a$ and sends to Bob: 
        \begin{align*}
            A = g^a \mod p
        \end{align*}
    \item Bob also chooses a secret integer $b$ and sends to Alice:
        \begin{align*}
            B = g^b \mod p
        \end{align*}
    \item Alice calculates $B^a\mod p$.
    \item Bob calculates $A^b\mod p$.
    \item The result for both of them is the same: an integer $s$ between $1$ and $p - 1$, which is their shared key that they can use for encryption of their communication from now on.
\end{enumerate}
The reason why Alice and Bob arrive at the same value for $s$ is the following:
\begin{equation*}
    A^b \mod p = (g^a \mod p)^b \mod p = g^{ab} \mod p
\end{equation*}
\begin{equation*}
    B^a \mod p = (g^b \mod p)^a \mod p = g^{ba} \mod p
\end{equation*}
The equality holds, since $g^{ab} = g^{ba}$.

In the Diffie-Hellman protocol only the integers $a$ and $b$ must be kept secret. The shared key $s$ can be computed by both of them using their own secret numbers. Since all other numbers are exchanged over a public network, an eavesdropper could easily overhear them. However, in order to compute the secret between Alice and Bob, the adversary must be able to compute the discrete logarithm \cite{mccurley1990discrete}. Since there is no efficient way known yet in order to do this, it is assumed to be hard and thus, the Diffie-Hellman Key Exchange protocol is secure against eavesdropping.

However, the protocol is vulnerable to man-in-the-middle (MITM) attacks on the wireless medium. Since neither Alice, nor Bob authenticated themselves in any way, the adversary can impersonate them, hide their messages and send another instead. If Lucifer is the attacker, he could receive the message from Alice, hiding it from Bob, and send her back:
\begin{equation*}
    L = g^l \mod p
\end{equation*}
with $l$ the secret integer of Lucifer. Lucifer does the same with Bob - he hides his message from Alice and sends him the number L he sent to Alice. Now, Alice and Bob have established a secret key with the adversary instead of with each other. The data that they would exchange over the wireless medium from now on would not be private anymore since Lucifer would receive it and possibly manipulate it. 

Since the Diffie-Hellman Key Exchange protocol is also used in the Push-Button Configuration, the vulnerability exists there as well. It could be prevented only by authentication of the parties, which requires additional protocols (e.g., a modified Diffie-Hellman protocol with authentication) or an out-of-band channel. However, this cannot be used in PBC since, as discussed, the overhead that it brings with itself contradicts the main idea of PBC - simplicity.

\subsection{IEEE 802.11}\label{sec3.2}
Since both PBC and TEP are based on the IEEE 802.11 standard, we summarize the most relevant aspects of it that are required for the further understanding.

As one of the world's most widely used wireless computer network standards, IEEE 802.11 provides the basis for the so-called \textit{Wi-Fi} devices. The Wireless Local Area Network (WLAN) communication relies on the specified by IEEE 802.11 set of media access control (MAC) and physical layer (PHY) protocols \cite{crow1997ieee}. It is used in home and office networks to allow laptops, smartphones, printers, and other devices to communicate with each other and access the Internet without connecting wires.

IEEE 802.11 is based on the Carrier-Sense Multiple Access with Collision Avoidance (CSMA/CA) method. It is a popular method in computer networking, in which prior to transmitting, a node first listens to the shared medium to determine whether another node is transmitting or not (including non 802.11 devices). The standard requires end-to-end packets to be separated by an interval called the DCF Inter-Frame Spacing (DIFS), whose value varies depending on the version of IEEE 802.11 that is used – e.g., 28 $\mu s$, 34 $\mu s$ or 50 $\mu s$. If there is no energy on the wireless medium for the DIFS period, the node does not start transmitting immediately, but waits for a period of time. The number of time slots, also known as backoff-slots, that the node has to wait before transmission, is chosen randomly from the interval $\{0,1,2, ... , min \{2^{c+k-1}-1,255\}\}$, where $c$ is dependent from PHY (usually $c=4$) and $k$ is the number of the previous unsuccessful transmission attempts, like in Binary Exponential Backoff \cite{goodman1988stability}. Since $c > 0$ is a constant, there is always a so-called \textit{Contention window} before transmitting a frame. After this period of time, also known as \textit{Backoff interval}, the node starts transmitting its data. This backoff is needed to avoid collisions with high probability, given that the other honest 802.11 nodes also wait for the DIFS period after the medium has become idle.

In order to be marked as successful transmitted, a frame has to be confirmed, i.e., the receiver has to send an acknowledgement (ACK) to the sender that it has received the frame. To do so, the receiver also has to wait for the medium to become idle, but instead of waiting for the DIFS period afterwards, it can transmit the acknowledgement after a shorter duration of 10 $\mu s$, called the Short Inter-Frame Spacing (SIFS). Since in a wireless public channel there are always multiple nodes transmitting at the same time and collisions happen often, these acknowledgements are useful.

Furthermore, an optional mechanism called Request to Send/Clear to Send (RTS/CTS) can be used in order to reduce frame collisions. If a node has a message to transmit, it has to send a Request to Send (RTS) to a base station (e.g., to the wireless access point) and then to wait for an answer. The node is allowed to transmit only after receiving a Clear to Send (CTS) message from it. It means that the medium is idle and all of the honest nodes would remain silent for the duration of the node's transmission and can send a RTS only after this period of time. Thus, the medium is reserved for a defined time slot and the node can transmit its data. Collisions are avoided, however, not entirely prevented, since there is no guarantee that all of the other nodes would have received the CTS message for that node.

The RTS/CTS mechanism works also without a base station. If a device is in ad-hoc mode (i.e., a group of IEEE 802.11 devices communicating with each other) broadcasting of RTS/CTS messages is used instead. Moreover, even if a device does not belong to the same service set, it has to be able to process a received CTS message and behave as required.

Another mechanism, similar to the RTS/CTS principle, is the CTS-to-Self. It is particularly useful in mixed environments with devices that support different 802.11 versions. Due to the different modulations used, they do not understand each other and might end up transmitting a frame at the same time. In order to avoid this situation, a CTS-to-Self method is implemented. When a device wishes to transmit in the medium, it sends a CTS frame without RTS frame preceding it. This helps the other devices in the network to understand that there will be a transmission and will prompt them to be quiet and continue to wait to access the medium.

Honest 802.11 can send with the minimum bit rate of 1 Mbps. Since the maximum packet size allowed by the lower layers is typically 1500 bytes, an honest node can occupy the channel for a maximum of 12 ms.

\subsection{Push-Button Configuration}\label{sec3.3} 
The Push-Button Configuration was introduced by the Wi-Fi Alliance as part of the Wi-Fi Protected Setup (WPS) standard \cite{alliancewi}. It aims to deal with devices that do not carry the interface to enter a password or PIN and to ease the security setup process for ordinary users. 

In order to pair two devices (e.g., a printer and a wireless access point) according to the PBC standard, a user has to press a button on both devices within the so-called walk time, which is usually 120 s. The device that wants to connect to the network is called an \textit{enrollee}, whereas the device that can give credentials for joining the network - a \textit{registrar}.

Once the enrollee's button is pushed, it starts actively searching for a registrar for the duration of the walk time by sending probes on the available 802.11 channels. It requests a reply from registrars whose PBC button has also been pressed and if a reply is received, it is stored and the enrollee continues scanning the other IEEE 802.11 channels. At the end of the walk-period, if more than one registrars have sent a reply, an error will occur, indicating that the pairing was unsuccessful and the user needs to try again later. If only one registrar responded, the enrollee assumes that it is the correct one and proceeds with the registration protocol. A connection is established using the received Diffie-Hellman key from that reply.

On the other hand, pressing the button of the registrar causes it to first check how many enrollees tried to connect in the last two minutes (during the so-called monitor time), i.e., how many requests were sent on his channel. An error is given immediately and it is indicated that the user has to retry, if more than one enrollee was detected trying to connect. Otherwise, the registrar listens to enrollees trying to connect for the duration of the walk time. If only one enrollee tried to do it at the end of it, the registrar answers according to the specification. A shared key is established using the Diffie-Hellman protocol, and the credentials for the network are sent to the enrollee. Otherwise, an error is sent.

\subsection{Attacker model}\label{sec3.4}
Apart from the fact that using the Diffie-Hellman Key Exchange protocol protects against passive eavesdroppers, the security of PBC essentially relies on the needed physical access to the devices. Nevertheless, Gollakota et al. describe 3 possible scenarios, in which PBC is vulnerable to an active MITM attack:
\begin{enumerate} 
    \item \textbf{Collision}: An adversary can create a collision with the enrollee's message. Immediately after, he can send his own, impersonating the enrollee. The registrar will notice only one connection attempt, and will thus allow pairing.
    \item \textbf{Capture effect}: An attacker can transmit a message at a much higher power and jam the enrollee's connection attempt, so that the registrar would again catch only one signal.
    \item \textbf{Timing control}: An attacker can transmit his connection attempt and constantly occupy the medium afterwards, so that the enrollee does not get the chance to send his own message at all as an honest 802.11 node. Again, only one message would be received by the registrar and the pairing would be successful.
\end{enumerate}
All these three scenarios gain the adversary access to the network and the data being transmitted. Hence, the attacker could now intercept and alter any future messages between the enrollee and the registrar without being detected. 

Defending against such kind of attacks is the main contribution of TEP. It aims to cope with the problem of authenticating key exchange messages between two wireless devices, while an active adversary might be trying to mount a MITM attack. Thus, it is important to clarify what the threat model we have to deal with is, i.e., the assumptions about the adversary that we are securing the protocol against. The following assumptions about the attacker were introduced by Gollakota et al. and used in the TEP implementation \cite{266516}, as well as in the model checking performed on it: 
\begin{itemize}
    \item An attacker can eavesdrop on any message from the medium.
    \item An attacker can transmit with an arbitrary power at any time, and thus overpower any message and replace it with his own.
    \item An attacker may be able to simultaneously receive and transmit signals (e.g., by using a multi-antenna system).
    \item An attacker could use directional antennas to ensure that only one of the pairing devices can hear its transmissions.
    \item An attacker can know the exact channel between the pairing devices, an the channel from the pairing devices to the adversary.
    \item An attacker can be anywhere in the network and is free to move.
    \item There could be more than one attacker that could collude with each other.
\end{itemize}
Thus, an adversary has full control over the medium. However, it is also assumed that:
\begin{itemize}
    \item An attacker \underline{cannot} have physical control over the pairing devices or their surroundings, e.g., the attacker cannot place either of the two devices in a Faraday cage to shield all signals. 
    \item An attacker \underline{cannot} efficiently determine the preimage of a hash and, in general, break traditional cryptographic constructs such as collision-resistant hash function.
    \item The PBC buttons operate according to the PBC standard.
    \item The user performs the PBC pairing as prescribed in the standard, i.e., the user puts the two devices in range and then pushes the buttons on both devices within 120 seconds.
\end{itemize}

\subsection{Bit-balancing Algorithm}\label{sec3.5}

We will now present the \textit{Bit-balancing Algorithm}, introduced by \cite{266516}, which has an important role for detecting tampering in TEAs (see \ref{sec4a}).

The algorithm takes as input a bit sequence of length $N$ ($N \mod 2 = 0$), and returns a balanced bit sequence of length $N + 2 \lceil$log $N \rceil$, i.e., with equal number of zeros and ones. 
We define the following variables:
\begin{itemize}
    \item $S_{i} \gets$ \textit{the bit sequence after flipping the first $i$ bits}
    \item $D_{i} \gets$ \textit{difference between the number of ones and zeros in the bit sequence after flipping the first $i$ bits}
    \item $M \gets$ \textit{the Manchester encoding of a number} 
    \item $R \gets$ \textit{the bit-balanced result}
\end{itemize}

The algorithm works as follows:

\begin{algorithmic}
\STATE $i\gets 0$
\STATE $S_{i}=S_{0} \gets$ \textit{input bit sequence}
\STATE $D_{i}=D_{0}\gets$ \textit{difference between the number of ones and zeros in the input sequence}
\WHILE{$D_{i}\neq 0$}
  \STATE $i\gets i+1$
  \STATE Compute $S_{i}$ by flipping the $i^{th}$ bit of $S_{i-1}$.
  \STATE Compute $D_{i}$, where $D_{i} = D_{i-1} \pm 2$, depending on whether the $i^{th}$ bit is one or zero.
\ENDWHILE 
  \STATE Compute the Manchester encoding $M$ of $i$.
  \STATE $R \gets S_{i}  ^\frown M$  
\end{algorithmic}

In each iteration, the algorithm checks the difference between zeros and ones in the bit sequence and if it is not zero, one bit is inverted. Hence, at the $i^{th}$ iteration, the first $i$ bits are flipped. This process is repeated until $D_{i}$ becomes zero. 

$D_{i} = 0$ is always reached with $i < N$, given $N$ is finite, due to the following reason. Since $N$ is required to be even, $D_{0}$ is also always an even number. When a bit is inverted, $D$ increases or decreases by 2, i.e., it remains even. If all $N$ bits of the initial bit sequence are flipped, then $D_{N} = -D_{0}$. However, zero is even and lies between $D_{N}$ and $-D_{0}$, which implies that this case is impossible because the algorithm would have already terminated.

The algorithm returns the input bit sequence with the first $i$ bits flipped, concatenated with the binary representation of $i-1$ in Manchester encoding \cite{tanenbaum2002computer}. The binary representation of $i-1$ is filled with leading zeros to the length of $\lceil$log $N\rceil$ bits. The Manchester encoding turns a 0 bit into 01 (or 10, depending on the implementation), and a 1 bit into 10 (or 01). Thus, the resulted Manchester code is also bit-balanced and is 2$\lceil$log$N\rceil$ bits long. This results in a bit-balanced sequence of total length $N+2\lceil$log $N\rceil$. The mapping is injective, and thus the result can be decoded in linear time. In Table \ref{tab1}, an example run of the algorithm for the bit sequence 1000 of length $N = 4$ is shown.

\begin{table}[t]
\caption{Example run of the BBA}
\renewcommand{\arraystretch}{1.1}
\setlength{\tabcolsep}{\tabcolsep}
\begin{center}
\begin{tabular}{p{0.27\textwidth}p{0.1\textwidth}}
\textbf{Input sequence} $S_{0}: 1000$& $D_{0}=-2$ \\
\hline
$i=1: S_{0}=1000 \rightarrow 0000=S_{1}$& $D_{1}=-4$ \\
$i=2: S_{1}=0000 \rightarrow 0100=S_{2}$& $D_{2}=-2$ \\
$i=3: S_{2}=0100 \rightarrow 0110=S_{3}$& $D_{3}=0$ \\
\hline
Binary representation of $i-1=2$: $10$ \\
Manchester encoding of $0b10$: $M=1001$ \\
\hline
\textbf{Result:} $S_{3} ^\frown M = 01101001$ \\
\end{tabular}
\label{tab1}
\end{center}
\end{table}

\section{Tamper-Evident Pairing}\label{sec4} 
Tamper-Evident Pairing is a modified version of PBC that aims to prevent MITM attacks. It provides simple and secure wireless pairing that does not use any out-of-band channels. To achieve this, a new security primitive, named a tamper-evident announcement (TEA), has been designed. TEA is a unidirectional announcement protocol that can be send in both directions (enrollee to registrar and vice-versa). It guarantees that under the described attacker model an adversary \textbf{cannot}
\begin{itemize}
    \item tamper with the payload of a TEA message, and
    \item mask the fact that a TEA message has been even transmitted.
\end{itemize}
TEP uses TEA for exchanging the Diffie-Hellman public keys between the PBC enrollee and registrar.


TEP is based on the fact that Wi-Fi devices can not only receive packets, but also simply measure the energy on the channel, which is part of the 802.11 standard requirements. This provides the opportunity to encode a bit of information as a time-slot where energy is present or absent on the wireless medium. Under the assumption that an attacker does not have the ability to remove energy from the medium, this means that an attacker cannot turn an on-slot into an off-slot. The needed terminology for understanding how the modified protocol of PBC works in detail as in \cite{266516} is given in Table \ref{tab2}.
 
\begin{table}[t]
\caption{Terms used to describe TEP \cite{266516}}
\renewcommand{\arraystretch}{1.1}
\setlength{\tabcolsep}{\tabcolsep}
\begin{center}
\begin{tabular}{p{0.1\textwidth}p{0.33\textwidth}}
\textbf{Term}&{\textbf{Definition}}\\
\hline
Tamper-evident announcement& A wireless message whose presence and the integrity of its payload are guaranteed to be detected by every receiver within radio range. \\
\hline
Synchronization packet& An exceptionally long packet whose presence indicates a TEA. To detect a synchronization packet, it is sufficient to detect that the medium is continuously occupied for the duration of the synchronization packet, which is 19 ms. \\
\hline
Payload packet& The part of a TEA containing the data payload (e.g., a device public key) \\
\hline
On/off-slot& The interval used to convey one bit from sender to receiver. The slot time is 40 $\mu s$. The bits in the slots are balanced. \\
\hline
Occupied/on-slot& A slot during which the medium is busy with a transmission. \\
\hline
Silent/off-slot& A slot during which the medium is idle. \\
\hline
Sensing window& The interval over which the receiver collects aggregate information for whether the medium is occupied or silent. \\
\hline
Fractional occupancy& The fraction of time the medium was busy during a sensing window. \\
\end{tabular}
\label{tab2}
\end{center}
\end{table}

\subsubsection{Tamper-Evident Announcement}\label{sec4a}

\begin{figure}[t]
\centerline{\includegraphics[width=0.4\textwidth]{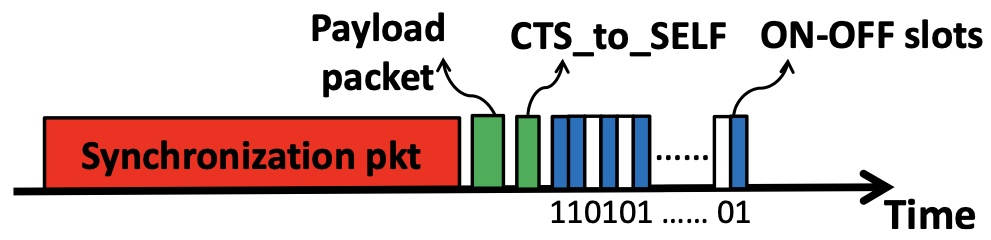}}
\caption{The format of a tamper-evident announcement (TEA). \cite{266516}}
\label{fig1}
\end{figure}

The goal of a TEA message is that an attacker can neither hide the transmitted message, nor modify its content without being detected. To achieve this, Gollakota et al. introduced a specific structure, given in Fig. \ref{fig1}. In order to get a clear idea of how TEP behaves when implemented on off-the-shelf devices, for each part of it there are details given such as duration or packet size. All of them are based on the implementation of TEP on Atheros AR5001X+ chipsets, provided by its authors.

First, a so-called \textbf{\textit{synchronization packet}} is transmitted. This is an exceptionally long packet, that indicates the start of a TEA. Normally, there is a maximum packet size that is permissible for the specific medium and if the packet is longer, it gets dropped by the receiver. However, the receiver does not need to decode the contents of a received synchronization packet, but only to detect the long burst of energy. That is why the maximum-sized packet allowed by the used hardware is transmitted. In the case of the Atheros AR5001X+ chipsets, it is 2400 bytes. Moreover, the synchronization packet needs to be transmitted at the lowest and most robust bit rate possible – 1 Mbps, in order to maximize the probability that all devices will receive it. So, the total duration of the transmission lasts $\frac{2400*8 bit}{1*10^6 bit/s} = 19,2 ms \approx 19 ms$. 

Although in practice it is very hard to cancel a signal in flight, theoretically an attacker that knows the exact transmitted signal and the channels to the receiver (which, according to the assumed attacker model, is possible), could also construct a signal that would cancel out the original signal at the receiver. To prevent this, the synchronization packet is filled with random data. Since this transmitted packet is exceptionally longer than any other transmission, if the receiver detects a burst of energy on the medium of at least its length, it uniquely identifies the start of a TEA. The synchronization packet also eliminates the opportunities of the adversary to hide the message by creating a collision, because such a long collision is rare, and thus easily detectable.

The TEA sender respects carrier-sense in the medium access control (MAC) protocol, and waits until the medium is idle before starting transmitting its message. However, if the message cannot be transmitted by a specific time $t$, it is assumed that an adversary is trying to mount an MITM attack with timing control, i.e., the attacker is constantly occupying the medium in order to prohibit the enrollee from sending his message. Therefore, the sender overrides the MAC's carrier-sense and transmits the announcement anyway, so that the recipient detects the tampering attempt. This way, the problem with hogging the medium is also solved.

The synchronization packet is followed by the \textbf{\textit{payload packet}}. The payload of a TEA contains the Diffie-Hellman public key of the sender that has to be exchanged. It has no restrictions on its content or encoding other than that it has to be of fixed length. This aims to ensure that an adversary cannot truncate or extend the payload during transmission.

As shown in Fig. 1, there is a gap between the synchronization and payload packets. If it is large, other 802.11 nodes would sense an idle wireless medium, and start transmitting, thus appearing to tamper with the TEA. To avoid this, the TEA sender has to send the payload packet immediately after the synchronization packet with a maximum gap of a SIFS, which is much less than DIFS. This solves the possible problem since 802.11 honest nodes are only allowed to transmit if they find the medium continuously idle for a DIFS.

After the payload, a \textbf{\textit{CTS\_to\_Self packet}} is sent. This is the message defined in the 802.11 specification that informs the nodes that the sender of the CTS\_to\_Self is about to transmit data, and requests all other Wi-Fi devices to refrain from transmitting during a certain time period, in our case the time needed for the remainder of the TEA. Therefore, the medium is reserved for the specific TEA message and all honest nodes will remain silent. Since it is important that all devices can decode the CTS\_to\_Self message, similarly to the synchronization packet, it is transmitted at 1 Mbps bit rate.

Finally, a sequence of equally sized packets called \textbf{\textit{on/off-slots}} is sent. The on/off-slots are used to detect any tampering with the TEA payload. A slot can either be an on-slot – transmission of random data (similarly to the synchronization packet), or an off-slot – completely silent. A transmission slot is interpreted as a 1 bit, whereas a silent slot as a 0 bit. The data encoded in the on/off slots is shown in Fig. 2.

\begin{figure}[t]
\centerline{\includegraphics[width=0.3\textwidth]{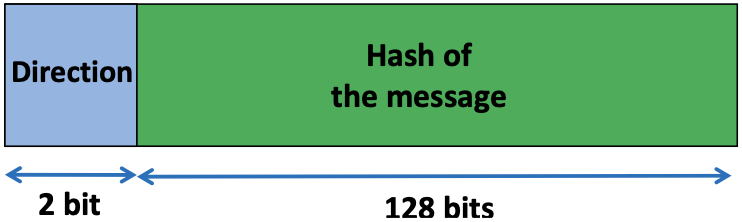}}
\caption{Data encoded in the on/off-slots. The first two bits specify the direction of the TEA. The following 128 bits contain the cryptographic hash of the payload. \cite{266516}}
\label{fig2}
\end{figure}

The first 2 bits encode the direction flag of the message, which specifies if this TEA message is a \textit{TEA request} by an enrollee, or a \textit{TEA reply} by a registrar. In the first case, the value is "10", and in the latter – "01". The remaining slots contain a cryptographic hash of the payload with length 128 bits. Protecting the encoded hash of the payload instead of the whole payload is efficient and suffices, since the hash function is injective. 

The hash slots are intended to prevent an attacker from modifying a message by transmitting simultaneously but with a higher power. Since an attacker cannot remove messages from the medium, he cannot turn a transmission slot into a silent slot. Nevertheless, the attacker can turn a silent slot into a transmission slot by transmitting data in this time interval, and thus change the hash. To detect tampering, TEA encodes the slots in a way that guarantees that exactly half of them are silent. A specially crafted bit-balancing algorithm (see \ref{sec3.5}) is applied to the 128-bit hash, prolonging it to 142 bits (with exactly 71 zeros and 71 ones). By using a bit-balanced hash, when an off-slot is changed into an on-slot, the balance between on- and off-slots is destroyed, making the hash invalid and the tampering detectable.
The CTS\_to\_Self makes sure that no honest node will change a silent slot into a transmission slot. The two direction bits precede the 142-bit bit-balanced hash, so, in total, 144 slots are sent. 

Since the only important information that a slot brings is the presence or absence of energy, they are transmitted at the highest bit rate in order to reduce the transmission duration. For IEEE 802.11a/b/g this is 54 Mbps, which corresponds to 40 $\mu$s for a single on/off-slot. It is allowed to assume that a slot needs exactly 40 $\mu$s due to the following changes in the 802.11 properties. The binary exponential backoff is disabled by setting the Contention window to 1 to ensure transmitting the slots at precise boundaries without delay. Moreover, the slot packets are put in the high-priority queue and the noise floor register is set to "high" to prevent carrier-sense backoff. 

According to the 802.11 standard, a single CTS\_to\_Self can reserve the channel up to 32 ms. The TEA message requires 144 slots. With a slot duration of 40 $\mu s$, this translates to about 5.8 ms which is less than the 32 ms allowed by the CTS\_to\_Self.

The CTS\_to\_Self message plays an important role here. Since the sender does not transmit during the off-slots, another 802.11 node could sense the wireless medium to be idle for more than a DIFS period and start transmitting its own packet during these off-slots. However, the 802.11 standard requires 802.11 nodes that hear a CTS\_to\_Self on the channel to abstain from transmitting for the period mentioned in that packet, which will ensure that no legitimate transmission overlaps with the slots. 

On the receiver side, the 144 slots are received by measuring the energy on the wireless medium. Hence, the receiver needs to distinguish noise from an actual transmission. To do so, the noise floor has to be set accordingly. There is a trade-off between the noise floor and the permissible distance between the pairing devices. Large distances imply weak signal, and therefore, the need of low noise floor to ensure detection. On the other hand, pairing devices that are close to each other allows a higher noise floor. In \cite{266516}, the default noise floor of -90dB is used, however, this could be modified.

Another important aspect for the receiver is how many and how long the sensing windows are, i.e., how often and how long the energy on the medium is measured. In order to minimize the overhead, the TEA receiver uses a long sensing window of 2 ms, until it detects a synchronization packet (a burst of energy longer than 17 ms). At this point the receiver switches to a 20 $\mu$s sensing window to accurately measure energy during slots. This corresponds on average to two measurements per one slot. The reason why the length of a sensing window is half the slot-length is that the synchronization between sender and receiver is not perfect. If the sensing windows are not aligned with the slots, up to half of the slot length will be measured in the wrong sensing window. By using 20 $\mu s$ sensing windows, at least one of the two sensing windows for a slot will be entirely in that slot, i.e., will not cross a slot boundary.

In the implementation on Atheros AP5001X+, proposed by Gollakota et al., to detect whether a slot was a transmission slot or a silent slot, the registers AR5K\_PROFCNT\_CYCLE and AR5K\_PROFCNT\_RXCLR are used. At the start of every sensing window, both of them get reset. The first one is incremented on every clock cycle/measurement and at the end of a sensing window has the number of total measurements $m$. The second one is incremented only if energy on the medium was detected and finally has the value $e$. 
For each sensing window, the \textit{fractional occupancy}, given by \begin{equation*}
    \frac{AR5K\_PROFCNT\_RXCLR}{AR5K\_PROFCNT\_CYCLE} = \frac{e}{m}
\end{equation*} is calculated and stored. It indicates the occupancy of the wireless medium during that sensing window and if the value is above a certain threshold (which is not specified by Gollakota et al.), the medium is considered occupied during the whole sensing window. 
Afterwards, the receiver determines which sensing windows per slot are the ones entirely in a slot – either all the even or all the odd. For this purpose, the variance of the fractional occupancy values is computed for the even and the odd windows. The sensing windows that are balanced and entirely in a slot have the higher variance, since overlapping would decrease it.

\begin{figure}[t]
\centerline{\includegraphics[width=0.5\textwidth]{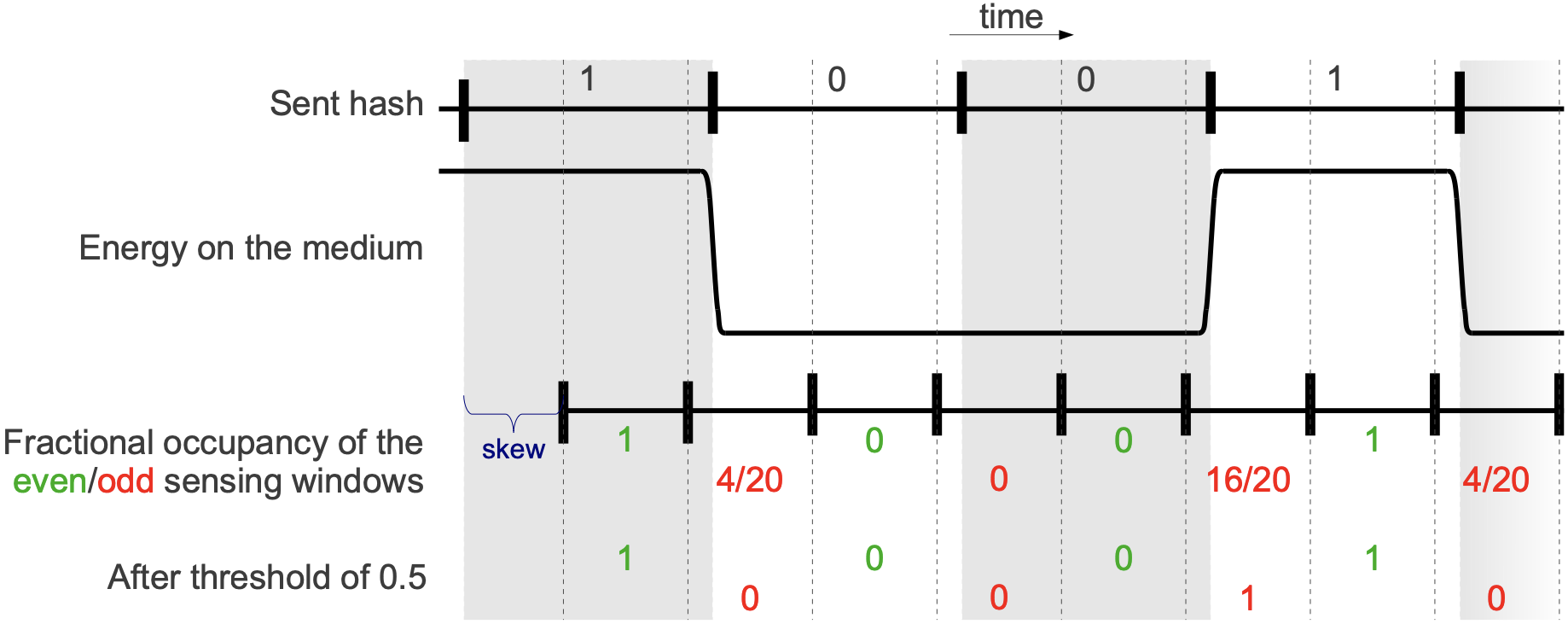}}
\caption{Sending and receiving the slots of a 4-bit hash. The even sensing windows are balanced and have higher variance. Therefore, they are chosen for the received hash. \cite{kersten2013using}}
\label{fig3}
\end{figure}

Fig. 3 illustrates this process for 4 transmitted bits and threshold, with a skew that represents the imperfect synchronization of sender and receiver. After all measurements are done, the fractional occupancy for each sensing window is calculated. Since the measurements are also not always perfect, the fractional occupancy is compared to a threshold value (here 0.5) to determine which slot was occupied. Finally, the even sensing windows are chosen, since they are balanced and with higher variance. Hence, they must fall entirely within the 40 $\mu s$ slots. The example shows that the received bit sequence is equal to the transmitted.

In case of a TEA message from an enrollee to a potential registrar, the CTS\_to\_Self packet additionally reserves the medium for a DIFS period past its slot transmissions. This serves two purposes. On the one hand, the registrar can immediately reply with its own TEA message and legitimate nodes cannot hog the medium and delay the registrar's response. On the other hand, it avoids under-utilization of the wireless medium which would happen if the channel is reserved for the entire length of a TEA reply and no registrar is present. Hence, a present PBC-activated registrar $must$ start transmitting its response message within the DIFS period. Otherwise, other legitimate devices will resume transmissions promptly. 

\subsubsection{Detecting tampering with TEA}\label{sec4b}

To determine if a TEA message has been tampered with, a TEA receiver performs several steps. 
First, it conservatively interprets any burst of energy during the monitor time at least as long as the synchronization packet as the start of a TEA. After receiving a sync packet, the receiver expects a payload and on/off-slots. If the receiver does not receive them (e.g., simply because the detected burst of energy was not a sync packet at all), a missed message is signaled anyway. However, since the synchronization packet is longer than any regular Wi-Fi transmission, the probability of these false positives is minimized.
Furthermore, this guarantees that the adversary cannot tamper with the presence of a TEA by masking it out, since, according to the attacker model, he cannot eliminate power on the medium even if he knows the exact communication channel and is fully synchronized with the transmitter.

Afterwards, the TEA receiver decodes the payload and the hash bits encoded in the on/off-slots and verifies them. If the decoding is not successful, the hash bits from the on/off-slots are not bit-balanced or they do not match the payload's hash, the receiver conservatively indicates tampering and aborts the pairing process. Once tampering is detected, the user is required to retry with the pairing later.  

As Gollakota et al. proved, a node would never miss TEA messages, even if they overlap with its own transmission.
In the case of a standard 802.11 packet transmission, the node samples the medium before and after its transmission. Since a standard transmission sent at the lowest rate lasts about 4 ms and the synchronization packet – 19 ms, it is unlikely to miss even if the entire transmission overlapped with part of a synchronization packet.
In case of a transmission of a TEA request/reply, there are several checks needed in order to ensure detection.
First, the direction bits are used to detect if there is a TEA transmission that is perfectly synchronized with this one, i.e., both are entirely overlapping. Since the direction flag for a request is "10" and for a reply "01", there is always one off-slot. The node checks the energy level on the medium during that slot and if the medium is not idle, an overlap is returned.
Additionally, the node samples the medium before and after every sync packet, and after the last on/off-slot. If energy is detected, it assumes that it may have missed a partially overlapping TEA and returns overlap. The total length of the on/off-slots is shorter than the length of the sync packet, so sampling the medium after the end of the sync packet (i.e., before the start of the payload and slots) and after the end of the slots suffices to detect a partially overlapping sync packet. This is the reason for the gaps in Fig. \ref{fig1} between the different parts of a TEA.

\subsubsection{Tamper-Evident Pairing}\label{sec4c}

Based on TEA, we will now describe how TEP – a modified version of the PBC protocol – avoids man-in-the-middle attacks.

As in the current PBC protocol, the enrollee scans the 802.11 channels repeatedly in round robin manner, once its button is pressed. At first, it respects carrier-sense, but if the medium is continuously occupied for a given time frame (e.g., 1 s), it overrides it and transmits its TEA request anyway (a TEA message with its public key and direction flag "10"). Afterwards, the enrollee waits for a TEA response from a registrar to store, who is required to immediately respond. After a specified period of time, it moves to the next 802.11 channel and repeats the process till the end of the PBC's walk time period. 

The registrar follows a similar protocol. Once the PBC button is pressed, it starts listening for possible TEA requests on its 802.11 channel. Every time a TEA request is received, the registrar records the message, payload and immediately sends its own TEA message in response, containing its public key. It is safe to reply immediately because the sender's TEA message contained a CTS\_to\_Self, which reserved the medium for the registrar's reply.

After the end of the PBC's walk time, both the enrollee and the registrar check the stored TEA replies. If they both received exactly one unique public key and no tampering with the TEA messages was detected, TEP proceeds with the pairing. In each other case, an error is signaled that either more than one pair of devices attempt to pair, or an adversary is mounting an attack. Either way, it is required for the user to retry.

\subsubsection{Optimization} 
The above described extension of PBC to use TEA requires both the enrollee and registrar to wait for full 120 s before completing the pairing. This is the only way to guarantee that at the end of the searching process both will have received the other party's response. The reason is that the devices do not know when exactly the buttons are pressed, just the period of time. To eliminate this delay, Gollakota et al. proposed an optimization that does not interfere with the security properties, but makes the pairing significantly faster.

The user needs to push the button on the enrollee first, afterwards the button on the registrar and lastly, once again the one on the enrollee. In this approach, the registrar will know that the enrollee's button has been already pressed and needs to wait just for the time it takes for the enrollee to cycle through all of the 802.11's channels, which, according to \cite{266516}, is less than 12 s. On the other hand, the second push of the enrollee's button eliminates its waiting time. Since the enrollee knows that the registrar is active, it just needs to cycle once through the 802.11 channels to find the registrar. This approach does not have any effects on the security and the worst case pairing time is as before, if the user does not push the enrollee's button once again in 120 s time period.

\section{Security Evaluation of TEP}
\subsection{Security proof by Gollakota et al.}
The resistance of TEP against MITM attacks for the presented in \ref{sec3.4} attacker model, is shown manually by its authors, i.e., no theorem prover tools or model checking are used. First, the used concepts are defined formally and based on these definitions, consequently is proven that TEA as well as TEP are tamper-resistant and thus, secure to MITM attacks.

According to Gollakota et al., a message is \textbf{\textit{tamper-evident}} \textit{if and only if its content cannot be changed without detection at the receiving party and its transmission cannot be hidden from the other party}. TEA is based on the assumption that an adversary cannot cancel out energy on the medium even if it knows the exact communication channel between two devices. However, this assumption is valid only if the transmitted signals are random, and hence unpredictable. Gollakota et al. proved by contradiction that the statement in that case always holds. Moreover, it implies that an adversary cannot hide the transmission of the synchronization packet, nor turn an on-slot into an off-slot, and thus change the hash of the payload.
Using this and given the transmitter and receiver are within range and able to sense the medium, it is proven that TEA is tamper-evident. The statement was proven to hold even if the receiver transmits its own messages at the same time. 
Lastly, TEP's security is shown based on the already proved properties. Since all of the theoretically possible cases are covered, the presented proof suffices and TEP is indeed secure against MITM attacks.

\subsection{\textsc{Uppaal}}
One year later, in 2012, a simple model of TEA and TEP in \textsc{Uppaal} was made by M. Drijvers \cite{drijvers2012model}. \textsc{Uppaal} is a tool with which properties about real-time systems, modeled as networks of timed automata, can be verified \cite{bengtsson1995uppaal}. Each network, also called a template, represents the different aspects of the model, i.e., the parts of it that are mostly independent from each other. A template has its own variables and function declarations.

\subsubsection{TEA Model}\label{5B1}
The TEA model in \textsc{Uppaal} consists of four templates: \textbf{sender}, \textbf{receiver}, \textbf{wireless medium}, and \textbf{adversary}. We will briefly explain their implementation ideas and more specific, the restrictions that were imposed for different reasons. The code and the implementation itself could be found in \cite{drijvers2012model}.

One of the templates of the TEA model is the \textbf{sender}. Before sending a TEA, the sender needs to know the payload, compute the payload's hash and bit-balance it.
Since calculating the hash and the bit balancing are tasks that both the sender and the receiver do internally, the communication between them does not matter. Therefore, both functions do not have an influence over the security properties of TEA, and more specifically over its resistance against MITM attacks. The bit-balancing algorithm, however, is implemented in the template as a user-defined function as specified.
In order to reduce the number of possibilities and the model-checking time, the payload has a specific value and is considered as given, even though it is supposed to be chosen nondeterministically. 

The \textbf{wireless medium} is implemented as specified in the TEP protocol. A sender or adversary can send data by storing it in a global variable. The adversary may additionally edit the sender's data before transmitting it to the others, exactly like according to the attacker model. Moreover, a boolean variable is used to indicate if the wireless medium is busy or idle.

The next template in the TEA model is the \textbf{adversary} with its ability to change the content of any message. When the wireless medium receives a message to send, it lets the adversary change its payload. Contrary to the attacker model, however, it can only be changed to a predefined value. Drijvers restricts the model in such a way in order to reduce the model-checking complexity due to high number of possibilities. Nevertheless, he argues that this limitation does not prevent attacks, because only the fact that the content is changed is what matters for MITM attacks, not the specific value.

According to the attacker model, the adversary can also transmit data in any moment, especially during the on/off-slots. In the \textsc{Uppaal} model, however, he is limited to transmitting only in predefined time frames for every on/off-slot, which aims to keep the model computationally tractable. Hence, for each 40 $\mu s$ slot, the adversary in the model is allowed to transmit during the first half of the slot, during the second half, during the whole slot, or not at all. The reason why this is acceptable is that for a sensing the medium node there is no difference between one and more than one nodes transmitting (simultaneously), i.e., what matters is whether there is energy on it or not, and the adversary has still the opportunities to perform tampering. Whenever this happens, a boolean variable $tampered$ is set to true.

The \textbf{receiver} waits for a synchronization packet and afterwards, for a payload and on/off-slots to store. Like in the original TEA implementation, 20 $\mu s$ sensing windows for the on/off-slots are used, in which the medium is checked repeatedly whether it is idle or busy. At the end of a sensing window, the fractional occupancy is stored. Nevertheless, the value is approximated because \textsc{Uppaal} does not support float variables. The fractional occupancy is first multiplied by a constant $c \in \mathbb{N}$ and then rounded by means of a user-defined function. When computing and comparing the variance of the even and odd slots, the factor $c$ does not need to be removed because $Var(c.X) = c^2. Var(X)$ and
     $ Var(c.X) < Var(c.Y) \leftrightarrow Var(X) < Var(Y)$.
After choosing the slots with higher variance and storing them in a boolean array, all elements of the array are compared to the $threshold$ of 0.5. Finally, the resulting bit sequence is compared to the balanced hash of the received payload and so it is decided if the received TEA has been tampered with. 

\subsubsection{TEP Model}
The TEP model consists of four templates as well: \textbf{user}, \textbf{enrollee}, \textbf{registrar}, and \textbf{adversary}. It assumes that the TEAs work as intended. Now, we will briefly explain the implementation ideas of the four templates.

The \textbf{user} starts the TEP by pressing either the registrar's or the enrollee's button. Within walk time, he is required to push the other button as well, so that the prerequisites for TEP are fulfilled.

The \textbf{enrollee}'s and \textbf{registrar}'s template are similar to each other. Moreover, they behave exactly as specified by Gollakota et al. when their buttons are pressed. The transmission of requests and listening to responses is realized by shared variables and broadcast channels. 
Since the model assumes that TEA work as intended, tampering with a TEA is expected to always be detected. This assumption is realized by using the mentioned boolean \textit{tampered} of a TEA, indicating whether this message has been tampered with. After the end of the walk time, both the enrollee and the registrar perform the usual checks before proceeding with the pairing. If during the walk time zero or more than one devices willing to pair were detected, an error is signaled. Otherwise, the boolean \textit{tampered} from that one received TEA is checked and the pairing is marked as successful only if this boolean is set to "false". Thus, the TEP protocol is exactly followed.

In the implementation of the \textbf{adversary}, there is a boolean variable for each its capability. Thus, they can be turned on and off independently and combined in different ways. The adversary can edit the payload of the transmitted messages. Whenever this happens, the TEA's boolean \textit{tampered} is set to true. Furthermore, the attacker can impersonate an enrollee or registrar by sending out probe requests/responses itself, using the same communication channel.

\subsubsection{Results}
The following desired properties have been checked on the TEA model:
\begin{itemize}
    \item The receiver always detects a transmitted TEA.
    \item Without an adversary interfering, the TEA is always delivered as expected.
    \item Whenever an adversary tampers with the TEA, it will be detected by the receiver.
\end{itemize}
The respective queries in \textsc{Uppaal} could be find in \cite{drijvers2012model}. All three properties were shown to hold with every configuration of the adversary:
\begin{itemize}
    \item Inactive adversary
    \item An adversary, who may edit the content of messages but does not send its own messages
    \item An adversary, who may transmit during off-slots but does not edit the content of other messages
    \item A fully active adversary, who may edit the content of messages and transmit during off-slots
\end{itemize}

The TEP model assumes TEAs to be working as intended. For the TEP model, 2 properties have been checked:
\begin{itemize}
    \item Without an adversary, the enrollee and the registrar always pair.
    \item An enrollee and a registrar would never be tricked to pair, using the wrong key, i.e., would never attempt to pair with the adversary.
\end{itemize}
Again, both properties held, which means that an adversary cannot impersonate a registrar to an enrollee or an enrollee to a registrar. Moreover, in neither case the pairing devices can be fooled and MITM attacks are therefore excluded. 

Nevertheless, the following (additional) restrictions that are imposed on the model and could have influenced the results have to be taken into account:
\begin{itemize}
    \item As a 32-bit application, the \textsc{Uppaal} verifier could only address 4 GiB of memory. Therefore, the TEA model has been adjusted so that the memory usage could be kept under the limit, e.g., a hash of length only 16 bits is used in order to prevent exceeding the memory.
    \item The attacker cannot transmit whenever he wants, but only in specified time frames.
    \item The fractional occupancies are rounded after multiplying with a factor and stored as integers due to the lack of float support in \textsc{Uppaal}. The chosen factor is also relatively small in order to avoid overflowing integers, because \textsc{Uppaal} does not support unsigned integers and longs.
    \item The TEA and TEP models are split. Testing them individually might have prevented some vulnerabilities.
\end{itemize}

\subsection{\textsc{Spin}}
In 2013, R. Kersten et al. investigated the TEP protocol further \cite{kersten2013using}. They focused on the fact that some parameters in the TEA protocol had not been fully specified by Gollakota et al. and tried to find out whether the values chosen for them would affect the TEP's correctness. For this purpose, TEA was modeled in the \textsc{Spin} model checker and numerous combinations of values were analyzed. 
The core idea of the authors was to check if an adversary with the same capabilities, as specified in \ref{sec3.4}, would be able to successfully change a payload packet's content and adapt the hash transmitted through the on/off-slots so that it matches the new payload's hash and is bit-balanced. However, since the tampering is detected through the bit-balanced hash and receiving it is the crucial moment, only its transmission is modeled in \textsc{Spin}. 

\subsubsection{Model Parameters}
We will briefly explain the implementation concepts for the parameters that are used in the model and play an important role for the model checking:

\paragraph{hash\_length}
This parameter indicates the length of the bit-balanced hash to send. The bit-balancing algorithm itself is not implemented. Instead, all possible bit-balanced hashes of this specific length are tested.

\paragraph{sw\_measurements}
This is the number of measurements per sensing window of length 20 $\mu s$. Normally, this number would depend on the hardware on which the protocol is implemented and could be different for each sensing window. However, in this model, it is fixed in order to reduce the model complexity. The simplification is tolerable since it does not affect the adversary's capabilities in any way and could not prevent an attack.

During an on-slot, the sender has to put energy on the medium for the number of clock-ticks it takes to do $2*sw\_measurements$ (one slot has two sensing windows). However, \textsc{Spin} does not have a built-in clock, so a discrete clock is implemented, whose clock-tick corresponds to a measurement taken by the receiver. This is acceptable, because the sender and receiver interact only through updating/measuring the energy on the wireless medium at most once and once, respectively, per clock cycle. The exact scheduling between them and their execution speed are thus not crucial. Nevertheless, the sender always executes first after a clock-tick in order to avoid the situation where the receiver measures before the sender has transmitted.

\paragraph{threshold}
This is the threshold value in \cite{266516} that the fractional occupancy of a sensing window is being compared to in order to determine if it was an on- or an off-slot. Therefore, it is  crucial for the protocol. However, its value is not specified by Gollakota et al. 
In this model, since \textit{sw\_measurements} is constant, the fractional occupancy is not calculated. This allows the threshold not to be a value between 0 and 1, but a value between 0 and \textit{sw\_measurements}, with which the number of measurements in which there was energy on the medium is compared.

\paragraph{skew}
As stated by Gollakota et al., the sender and the receiver might not be perfectly synchronized. This motivated the use of two sensing windows per slot which limits the duration of a possible clock skew up to 10 $\mu s$ ($\frac{1}{2}$ of the sensing window's length) and guarantees that either the even or the odd windows do not cross slot-boundaries.
Furthermore, a start of a TEA is recognised through detecting a continuous burst of energy for at least 19 ms. Kersten et al. argued that neither the maximum length of this energy burst is specified, nor the exact synchronization point. Therefore, the adversary could easily prolong it by putting energy on the wireless medium. The difference with the 19 ms introduces the additional clock skew. The parameter \textit{skew} stores the sum of all skews and is also measured in clock-ticks. It can be included in the model by letting one of the processes wait for a number of clock-ticks before starting with its tasks.

\subsubsection{Model Processes}
The model consists of the following four processes, which start after all possible hashes of the given length are non-deterministically generated:
\begin{itemize}
    \item \textbf{Clock}, which generates the clock-ticks.
    \item \textbf{Sender}, which transmits the generated hash bit by bit by putting energy on the medium or not. Between each two of them, it waits for $2 * sw\_measurements$ clock-ticks. At the end of the hash, it keeps clock-ticking, in case the receiver process is still running.
    \item \textbf{Receiver}, which waits for \textit{skew} clock-ticks before starting measuring the energy on the medium (once on each clock-tick). After measuring \textit{sw\_measurements} times, the corresponding bit is stored for that sensing window (1 if there were more than \textit{threshold} measurements with energy detected and 0, otherwise). After the measurements for all sensing windows are done, it is verified if the even or the odd sensing windows have an equal number of on- and off-slots.
    \item \textbf{Adversary}, which can increase the energy on the medium whenever he wants to.
\end{itemize}

\subsubsection{Results}
The model checking with \textsc{Spin} revealed a vulnerability of TEA, and thus of TEP. The following desired property was disproved: \textit{If the received hash is not equal to the sent hash, tampering is always detected}. Kersten et al. provided a scenario, which illustrates the found vulnerability.

\begin{figure}[t]
\centerline{\includegraphics[width=0.5\textwidth]{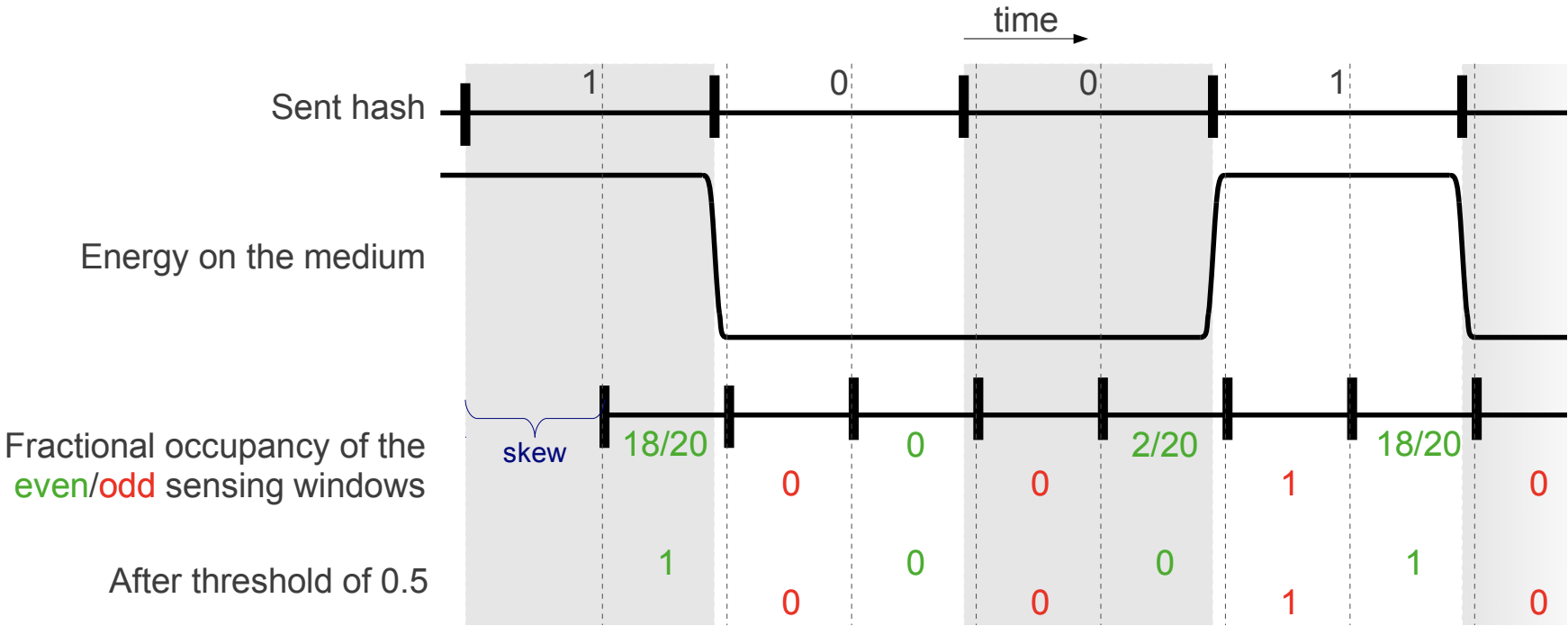}}
\caption{Scenario with no adversary, and thus correctly received hash of length 4, but in which TEP is vulnerable. \cite{kersten2013using}}
\label{fig4}
\end{figure}

Fig. \ref{fig4} shows the case with no active adversary. Since the even sensing windows are bit-balanced and have higher variance, they are chosen by the receiver. Hence, the received hash equals the sent one. Threshold of 0.5 implies the value of $\frac{1}{2}.sw\_measurements$.

\begin{figure}[t]
\centerline{\includegraphics[width=0.5\textwidth]{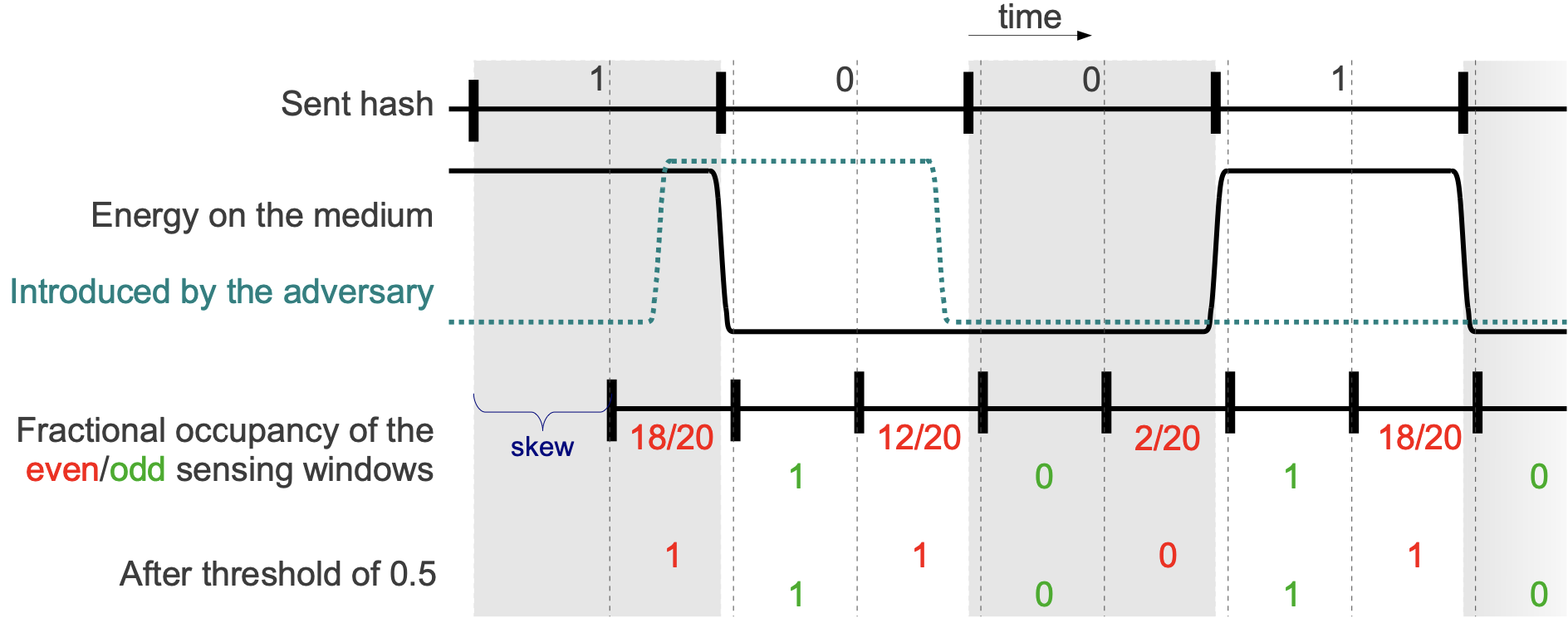}}
\caption{Scenario with an active adversary who successfully changes the received hash to 1010 without being detected. \cite{kersten2013using}}
\label{fig5}
\end{figure}

Fig. \ref{fig5} shows a successfully mounted attack in the same scenario. An active adversary introduces energy on the wireless medium for the shown time period. This tricks the receiver to choose the odd sensing windows and accept the hash 1010 without detecting tampering.

Kersten et al. showed that an adversary can turn any 1 bit in the hash into a 0, \textit{if and only if it is immediately followed by a 0}. Given that the hash is bit-balanced and consists of 50\% zeros and 50\% ones, this gives the adversary numerous opportunities. Furthermore, the circumstances under which this vulnerability occurs were also closely investigated. The following was confirmed:
\begin{itemize}
    \item The hash length has no influence on the occurrence of the vulnerability.
    \item The vulnerability occurs if the following predicate holds:
    \begin{equation}\label{predicate}
        skew \geq sw\_measurements - threshold
    \end{equation}
\end{itemize}

Thus, if the \textit{skew} is large enough, a part of the measurements of the even windows could be moved to the following odd sensing window, giving the adversary a chance to change the hash. The statement was shown to hold for all combinations of values for \textit{sw\_measurements}, \textit{threshold} and \textit{skew} from 1-10.

\section{Discussion}
After Gollakota et al. introduced the TEP protocol in 2011 as an improvement of PBC in terms of MITM attacks, its security properties were additionally tested with two model checkers, namely \textsc{Uppaal} in 2012 and \textsc{Spin} in 2013. Although during the \textsc{Uppaal} model checking no vulnerabilities were found, the \textsc{Spin} model revealed a case, in which the correctness of the protocol is no longer guaranteed. Furthermore, the vulnerability boundaries were explored and defined by testing different combinations of values for the specific parameters.

The essential condition for successfully mounting a MITM attack appeared to be the clock skew, i.e., the delay with which the receiver starts measuring the energy on the wireless medium. Since the \textsc{Uppaal} model did not consider it at all, the model checking could not reveal the vulnerability. Moreover, TEA and TEP were tested separately and the TEP model assumed that TEAs work as intended. In order to reduce the TEA model complexity, many simplifications and abstractions were made (discussed in \ref{5B1}), that also contribute to hide the vulnerability.
Nevertheless, the TEP model was modeled accurately and since it was tested independently from the TEA model, the results can be accepted as reliable. This is also the main contribution of the \textsc{Uppaal} model checking. Similar to the \textsc{Spin} model checking, one can implement/improve only the TEA model. If no vulnerabilities are found, this suffices to claim that this statement holds for the TEP protocol as well.

The found vulnerability appears only when predicate \ref{predicate} holds. It is thus avoidable, e.g., by not enabling a clock-skew more than a specific value, by lowering the threshold or increasing the number of measurements per sensing window. To the current moment, this is the only restriction needed in order to ensure TEP's security.

Apart from checking TEP's security properties, Gollakota et al. also evaluated TEP's accuracy using existing operating systems and off-the-shelf 802.11 hardware (Atheros chipsets with Ath5k driver). The most challenging and important part of TEA are the on/off-slots. Multiple experiments tested if a transmitter can accurately schedule their transmission and if a receiver can accurately distinguish them.
The results showed that the transmitter is able to schedule the on/off-slots at microsecond granularity. The median scheduling error for the bit sequence of alternating zeros and ones, requiring maximal scheduling precision, was less than 0,4 $\mu s$, which is sufficient.
Furthermore, the fractional occupancies make the on-slots perfectly distinguishable from the off-slots for a receiver within the sender's radio range. Thus, TEP is realizable on existing devices.

TEP's performance in operational networks was also evaluated by its authors. Due to cross traffic, false positives are possible, i.e., a node incorrectly declares that a TEA message has been tampered with, which unnecessarily delays the secure pairing.
In order to determine how often a false positive occurs due to mistaking a normal 802.11 packet for a TEA, it was first tested how often a sync packet is incorrectly detected. According to Gollakota et al., this could happen if either the legitimate traffic's bursts of energy last at least as long as the sync packet (19 ms), or if the TEP receiver cannot detect the DIFS/SIFS interval separating normal packets, and thus declares a stream of them as a single continuous energy.
However, the experiments in two operational networks with off-the-shelf hardware showed that both are very unlikely. All energy bursts lasted for less than 4.3 ms, whereby the duration of the majority of them was between 0.25 and 2 ms – significantly shorter than the sync packet's length. Furthermore, during the experiment no node confused a sequence of packets separated by DIFS as a single energy burst. 

A false positive also occurs if there are non-802.11 devices in range that may also transmit. Since they do not decode and respect the 802.11 CTS\_to\_Self packets, they could transmit during the on/off-slots and cause TEA nodes to declare tampering incorrectly. However, an experiment with Bluetooth devices in range showed that TEP is still able to perform a key exchange with only 1.4 attempts on average (and max. 4 attempts).

Therefore, TEP can accurately be implemented in existing Wi-Fi devices without the need of hardware changes and its performance is comparable with PBC's. Moreover, it protects against MITM attacks, when complied with the defined predicate, which makes the protocol a valuable improvement of PBC.

\section{Conclusion/Future work}
Tamper-Evident Pairing (TEP) is the first wireless pairing protocol that protects against MITM attacks without using out-of-band channels for authentication. TEP is built on a primitive, called Tamper-Evident Announcement (TEA), which guarantees that an adversary can neither tamper with the payload of a transmitted message without being detected, nor hide the fact that the message was sent in the first place. In this paper, we provide a comprehensive summary of everything known about TEP and include all of the security checks that were performed on it so far.

TEP's security properties against MITM attacks were first proven by its authors on paper. Afterwards, they were consequently tested with \textsc{Uppaal} and \textsc{Spin}. 

The \textsc{Uppaal} model checked the desired properties for TEA and TEP separately. All of the desired properties held for both models. In the absence of adversary, the enrollee and the registrar always pair, whereas in the presence of an adversary who is actively mounting a MITM attack, the tampering is always detected. 
However, the TEA model was significantly simplified in order to reduce the model complexity.

The \textsc{Spin} model addressed particularly the TEA's unspecified parameters in its specification. Based on the successfully proved security of TEP with \textsc{Uppaal}, only the TEA was model-checked. In particular, the sending of a hash through on/off-slots. Furthermore, a clock skew was included and a vulnerability was indeed found. This model checking proved to be very effective in both uncovering a vulnerability for specific values as well as in finding the condition indicating for which values the vulnerability is present.

This leaves room for future work. First, the TEP and TEA models could be merged into a single model because this might bring new vulnerabilities. In order to avoid any assumptions or simplifications and make the model as real as possible, a model checker with float support and enough memory is needed. Moreover, the opportunities for exploiting/avoiding the found vulnerability could be further investigated. Finally, a full formal proof by a theorem prover and a proof that the found vulnerability cannot occur when the defined predicate is not satisfied would be very useful. Since TEP can be implemented on off-the-shelf 802.11 devices and is practical in real world, it would be a huge advantage if it is fully explored and can be applied by the industry.

\bibliographystyle{ieeetr}
\bibliography{ref}

\begin{thebibliography}{10}

\bibitem{alliancewi}
W.-F. Alliance, ``Wi-{F}i {P}rotected {S}etup {S}pecification, version 1.0 h,
  2006,''

\bibitem{viehbock2011brute}
S.~Viehb{\"o}ck, ``Brute forcing {W}i-{F}i {P}rotected {S}etup,'' {\em Wi-Fi
  Protected Setup}, vol.~9, 2011.

\bibitem{journals/tit/DiffieH76}
W.~Diffie and M.~E. Hellman, ``New directions in cryptography.,'' {\em IEEE
  Trans. Inf. Theory}, vol.~22, no.~6, pp.~644--654, 1976.

\bibitem{kainda2009usability}
R.~Kainda, I.~Flechais, and A.~W. Roscoe, ``Usability and security of
  out-of-band channels in secure device pairing protocols,'' in {\em
  Proceedings of the 5th Symposium on Usable Privacy and Security}, pp.~1--12,
  2009.

\bibitem{266516}
S.~Gollakota, N.~Ahmed, N.~Zeldovich, and D.~Katabi, ``Secure {In-Band}
  {W}ireless {P}airing,'' in {\em 20th USENIX Security Symposium (USENIX
  Security 11)}, (San Francisco, CA), USENIX Association, Aug. 2011.

\bibitem{owre1992pvs}
S.~Owre, J.~M. Rushby, and N.~Shankar, ``{PVS}: A {P}rototype {V}erification
  {S}ystem,'' in {\em International Conference on Automated Deduction},
  pp.~748--752, Springer, 1992.

\bibitem{holzmann1997model}
G.~J. Holzmann, ``The model checker {SPIN},'' {\em IEEE Transactions on
  software engineering}, vol.~23, no.~5, pp.~279--295, 1997.

\bibitem{bengtsson1995uppaal}
J.~Bengtsson, K.~Larsen, F.~Larsson, P.~Pettersson, and W.~Yi, ``{UPPAAL} — a
  {T}ool {S}uite {F}or {A}utomatic {V}erification of {R}eal-{T}ime {S}ystems,''
  in {\em International hybrid systems workshop}, pp.~232--243, Springer, 1995.

\bibitem{cimatti1999nusmv}
A.~Cimatti, E.~Clarke, F.~Giunchiglia, and M.~Roveri, ``Nu{SMV}: A {N}ew
  {S}ymbolic {M}odel {V}erifier,'' in {\em International conference on computer
  aided verification}, pp.~495--499, Springer, 1999.

\bibitem{Basin2018}
D.~Basin, C.~Cremers, and C.~Meadows, {\em Model Checking Security Protocols},
  pp.~727--762.
\newblock Cham: Springer International Publishing, 2018.

\bibitem{drijvers2012model}
M.~Drijvers, M.~Van~Eekelen, and R.~Kersten, ``Model checking
  {T}amper-{E}vident {P}airing,'' 2012.

\bibitem{kersten2013using}
R.~Kersten, B.~v. Gastel, M.~Drijvers, S.~Smetsers, and M.~v. Eekelen, ``Using
  {M}odel-{C}hecking to {R}eveal a {V}ulnerability of {T}amper-{E}vident
  {P}airing,'' in {\em NASA Formal Methods Symposium}, pp.~63--77, Springer,
  2013.

\bibitem{6687314}
S.~Mirzadeh, H.~Cruickshank, and R.~Tafazolli, ``Secure {D}evice {P}airing: A
  {S}urvey,'' {\em IEEE Communications Surveys \& Tutorials}, vol.~16, no.~1,
  pp.~17--40, 2014.

\bibitem{kuo2007low}
C.~Kuo, J.~Walker, and A.~Perrig, ``Low-cost {M}anufacturing, {U}sability, and
  {S}ecurity: An {A}nalysis of {B}luetooth {S}imple {P}airing and {W}i-{F}i
  {P}rotected {S}etup,'' in {\em International Conference on Financial
  Cryptography and Data Security}, pp.~325--340, Springer, 2007.

\bibitem{4537388}
K.~M. Haataja and K.~Hypponen, ``{M}an-in-the-{M}iddle {A}ttacks on
  {B}luetooth: A {C}omparative {A}nalysis, a {N}ovel {A}ttack, and
  {C}ountermeasures,'' in {\em 2008 3rd International Symposium on
  Communications, Control and Signal Processing}, pp.~1096--1102, 2008.

\bibitem{4679061}
K.~Haataja and P.~Toivanen, ``Practical man-in-the-middle attacks against
  {B}luetooth {S}ecure {S}imple {P}airing,'' in {\em 2008 4th International
  Conference on Wireless Communications, Networking and Mobile Computing},
  pp.~1--5, 2008.

\bibitem{5374082}
K.~Haataja and P.~Toivanen, ``Two {P}ractical {M}an-in-the-{M}iddle {A}ttacks
  on {B}luetooth {S}ecure {S}imple {P}airing and {C}ountermeasures,'' {\em IEEE
  Transactions on Wireless Communications}, vol.~9, no.~1, pp.~384--392, 2010.

\bibitem{mccurley1990discrete}
K.~S. McCurley, ``The discrete logarithm problem,'' in {\em Proc. of Symp. in
  Applied Math}, vol.~42, pp.~49--74, USA, 1990.

\bibitem{crow1997ieee}
B.~P. Crow, I.~Widjaja, J.~G. Kim, and P.~T. Sakai, ``{IEEE} 802.11 {W}ireless
  {L}ocal {A}rea {N}etworks,'' {\em IEEE Communications magazine}, vol.~35,
  no.~9, pp.~116--126, 1997.

\bibitem{goodman1988stability}
J.~Goodman, A.~G. Greenberg, N.~Madras, and P.~March, ``Stability of {B}inary
  {E}xponential {B}ackoff,'' {\em Journal of the ACM (JACM)}, vol.~35, no.~3,
  pp.~579--602, 1988.

\bibitem{tanenbaum2002computer}
A.~S. Tanenbaum, {\em Computer networks, 4th edition}.
\newblock Prentice Hall, 2002.

\end{thebibliography}

\end{document}